# Transparent sub-diffraction optics: Nanoscale light confinement without metal

SAMAN JAHANI[1], ZUBIN JACOB[1*]

[1] *Department of Electrical and Computer Engineering, University of Alberta, Edmonton, Canada T6G 2V4*
*Corresponding author: zjacob@ualberta.ca*

**The integration of nanoscale electronics with conventional optical devices is restricted by the diffraction limit of light. Metals can confine light at the subwavelength scales needed, but they are lossy, while dielectric materials do not confine evanescent waves outside a waveguide or resonator, leading to cross talk between components. We introduce a paradigm shift in light confinement strategy and show that light can be confined below the diffraction limit using completely transparent artificial media (metamaterials with $\varepsilon_{ij} > 1$, $\mu_{ij} = 1$). Our approach relies on controlling the optical momentum of evanescent waves—an important electromagnetic property overlooked in photonic devices. For practical applications, we propose a class of waveguides using this approach that outperforms the cross-talk performance by 1 order of magnitude as compared to any existing photonic structure. Our work overcomes a critical stumbling block for nanophotonics by completely averting the use of metals and can impact electromagnetic devices from the visible to microwave frequency ranges.**

Modern computation and communication systems rely on the ability to route and transfer information using electronic and electromagnetic signals. Massive efforts over the last decade have been driven by miniaturization and integration of electronics and photonics on the same platform[1]. However, the diffraction limit of light is a fundamental barrier to interface micron scale waveguides to nanoscale electronic circuitry. Furthermore, dense photonic integration is hampered because crosstalk between waveguides increases as the separation between them is reduced.

At low frequencies, metals due to their high reflectivity can be used for confining light at the subwavelength scale[2]. At optical frequencies, metals can achieve the same task by coupling light to free electrons. This leads to a surface plasmon polariton (SPP) which shows properties of nanoscale waveguiding[3,4]. However due to absorption in metals this approach cannot guide light more than a few microns[3,5,6]. Furthermore, the dissipated energy leads to thermal issues which are especially significant in miniaturized circuits hindering dense photonic integration. Hence, low loss approaches to light confinement at the nanoscale are a fundamental necessity for photonics.

Prevalent all-dielectric nanophotonic approaches can be classified according to two fundamental principles governing them. One which utilizes the large index contrast between media to confine light within nanoscale slots[7-9] and the other which uses bragg reflection of waves in the bandgap of photonic crystals[10-12]. However, neither of the approaches functions on the evanescent fields outside the core of the resonator or waveguide. These unchecked evanescent waves are the fundamental origin of cross-talk in nanophotonics and this significantly limits the ability of these classes of dielectric waveguides for photonic integration[13,14].

In this paper, we surpass the diffraction limit of light by a new class of all-dielectric artificial materials that are lossless. This overcomes one of the fundamental challenges of light confinement in metamaterials and plasmonics: metallic loss. Our approach relies on controlling the optical momentum of evanescent waves as opposed to conventional photonic devices which manipulate propagating waves. This leads to a counterintuitive confinement strategy for electromagnetic waves across the entire spectrum. Finally, we propose a class of practically achievable waveguides based on these momentum transformations that exhibit dramatically reduced cross talk compared to any dielectric waveguide (slot, photonic crystal and conventional).

**Paradigm shift in light confinement strategy.** We introduce two distinct photonic design principles that can ideally lead to sub-diffraction light confinement without metal.

*i) Relaxed-Total Internal Reflection:* First, we revisit the conventional light confinement mechanism of total internal reflection that is widely utilized for waveguides and resonators. We consider a simple 2D case with an interface along the z-axis between medium 1 and 2 and TM polarized incident light. A habitual prejudice immediately leads us to conclude that $n_1 > n_2$ as a condition for total internal reflection of light moving from medium 1 to 2 (fig. 1.a). Here, we argue that the above is a sufficient but not necessary condition and the requirement can be relaxed to

$$n_1 > \sqrt{\varepsilon_x} \qquad (1)$$

where the z-axis is parallel to the interface and the x-axis is normal to it. Note that $\varepsilon_x$ is defined as the dielectric constant of medium 2 perpendicular to the interface. We call this condition relaxed-Total Internal Reflection. We provide a simple proof of this using momentum conservation of light parallel to the interface in uniaxial anisotropic media. The tangential momentum of light $k_z^{\parallel} = n_1 k_0 \sin\theta$ is conserved along the interface where $k_0 = \omega/c = 2\pi/\lambda$ is the free space wavevector of light and $\theta$ is the angle of incidence. Once the light enters the medium 2, even though the parallel momentum is conserved, the dispersion relation of TM polarized waves changes to

$$\frac{(k_z^{\parallel})^2}{\varepsilon_x} + \frac{(k_x^{\perp})^2}{\varepsilon_z} = (k_0)^2 \qquad (2)$$

We see that the perpendicular component of the wavevector in the second medium $k_x^{\perp}$ can be zero or imaginary (evanescent wave) if $k_z^{\parallel} > \sqrt{\varepsilon_x} k_0$ i.e. $n_1 > \sqrt{\varepsilon_x}$. Note that as expected, for angles of incidence greater than the critical angle of relaxed-TIR ( $\theta_c = \sin^{-1}(\sqrt{\varepsilon_x}/n_1)$ ) we have an evanescent wave decaying into medium 2.

*ii) Transforming optical momentum:* In conventional TIR, the evanescent wave penetrates considerably into medium 2. Here, we show how to transform the optical momentum of evanescent waves leading to a reduced penetration depth (skin depth) in medium 2 after TIR. The argument in the previous sub-section about relaxed-TIR opens up a fundamentally new degree of freedom for confining evanescent waves penetrating in medium 2: the component of the dielectric tensor parallel to the interface. The evanescent wave decay constant for TM polarized waves in medium 2 is given by

$$k_x^{\perp} = \sqrt{\frac{\varepsilon_z}{\varepsilon_x}} \sqrt{\varepsilon_x (k_0)^2 - (k_z^{\parallel})^2} \qquad (3)$$

The penetration depth of evanescent fields into the second medium is thus governed by the ratio of permittivity components $\sqrt{\varepsilon_z/\varepsilon_x}$. We thus arrive at the condition $\varepsilon_z \gg 1$ to increase the momentum of evanescent waves i.e. make them decay faster confining them very close to the interface. Note that since we have decoupled the total internal reflection criterion ($n_1 > \sqrt{\varepsilon_x}$) from the momentum confinement condition ($\varepsilon_z \gg 1$) both can be simultaneously achieved leading to a fundamentally new approach to light confinement in transparent media (fig. 1.b).

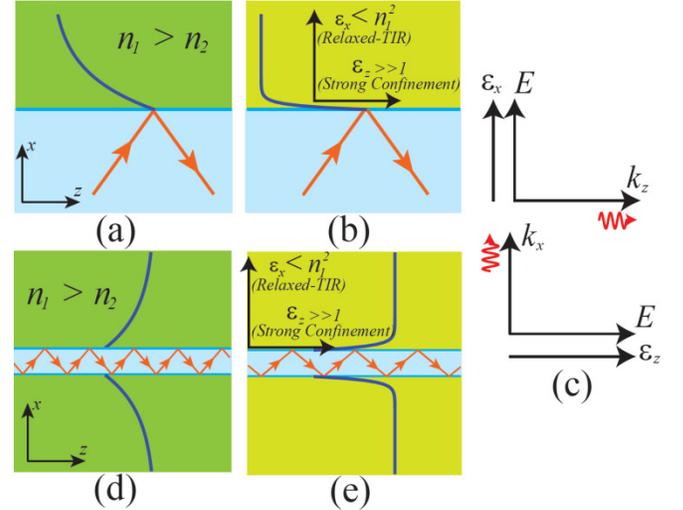

Fig.1. (a) Conventional total internal reflection: If $n_1 > n_2$ and the incident angle is larger than the critical angle, the light is totally reflected to the 1st medium and decays in the 2nd medium. (b) Relaxed total internal reflection: If $n_1 > \sqrt{\varepsilon_x}$ and the incident angle is larger than the critical angle, the light is totally reflected. However, the penetration depth can be decreased considerably if $\varepsilon_z \gg 1$. (c) Wave propagation along the optical axis of a uniaxial medium. As the electric field is perpendicular to the momentum direction, permittivity in specific direction controls the momentum in the perpendicular direction. (d) Conventional waveguide based on total internal reflection: As the core size is decreased, most of the power lies outside and decays slowly in the cladding. (e) Transformed cladding waveguide: Relaxed total internal reflection ($n_1 > \sqrt{\varepsilon_x}$) preserves the conventional waveguiding mechanism. Furthermore, the light decays fast in the cladding as the optical momentum in the cladding is transformed using anisotropy ($\varepsilon_z \gg 1$). Thus the wave can be confined inside the core giving rise to sub-diffraction optics with completely transparent media.

*(iii) Controlling optical momentum with dielectric anisotropy:* In essence, our non-resonant transparent

medium alters the momentum of light entering it. The upper limit to the momentum tangential to the interface is set by the dielectric constant perpendicular to the interface while the perpendicular momentum is increased by the dielectric constant parallel to the interface. This non-intuitive concept of controlling wave momentum in a given direction by the dielectric constant perpendicular to the phase propagation is depicted in figure 1(c.) It is seen that for plane wave propagation along the symmetry axes of anisotropic media, the field direction and the relevant dielectric tensor component are perpendicular to the wavevector. ($k_x$ governed by $\varepsilon_z$ and $k_z$ governed by $\varepsilon_x$).

**Sub-diffraction light confinement without metal.** We now show how the previous momentum transformations can be used for sub-diffraction confinement of light without metallic plasmons. Note that our approach can be applied to multiple devices across the visible, THz and microwave regimes but for the sake of elucidation, we consider a waveguide geometry.

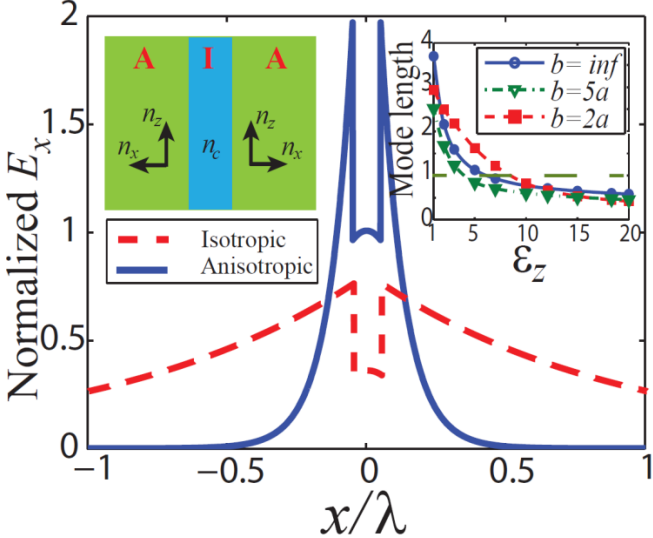

Fig. 2. Normalized tangential electric field of the TM mode for a glass slab waveguide with a size of $0.1\lambda$ surrounded with all-dielectric metamaterial cladding. The metamaterial has dielectric constants of ($\varepsilon_x = 1.1$ and $\varepsilon_z = 15$). On comparison with a conventional mode which has air as the surrounding medium, a rapid decay of the evanescent fields is observed. The plots are normalized to the same input electric energy. (inset) As the anisotropy of the cladding is increased the mode length decreases significantly below the diffraction limit with completely transparent media. This can be achieved with a cladding size (width b), three times that of the core (width a)

As the core size of a conventional slab waveguide is decreased, all modes are cut-off except the lowest order TE and TM modes. Even though these modes exist, most of the power actually lies outside the core and decays very slowly in air (cladding). We propose to use metamaterial claddings which transform the momentum of evanescent waves and confine the light within the core of the waveguide (Fig. 1 d and e). We call such waveguides as extreme skin depth (e-skid) waveguides. We emphasize the counter-intuitive nature of the waveguiding since the index of the cladding averaged over all directions is greater than that of the core. The relaxed-TIR condition still allows the lowest order mode to propagate similar in principle to the conventional case. Note that the confinement is achieved for the $TM_0$ mode since the dispersion relation is anisotropic only for TM waves.

In fig. 2, we show the field plots for the fundamental mode in a conventional waveguide and a 1D transformed cladding waveguide. This engineered anisotropy allows us to control the evanescent field outside the core. For the same input energy, we note the increased power in the core and striking difference between the evanescent decay in the cladding for the two waveguides.

We follow the conventional definition of mode length adopted from the concept of mode volume in quantum optics and widely used in nanoscale waveguide theory[5,15]. For the lowest order $TM_0$ waveguide mode it is given by

$$L_m = \int_{-\infty}^{\infty} W(x)dx / \max\{W(x)\},$$

where $W(x)$ is the energy density of the mode[5,15]. It is clear from the inset of Fig. 2 that the mode length is diffraction limited (above $\lambda/2n_{core}$) for conventional waveguides. However, once the cladding is made anisotropic, the mode length achieves sub-diffraction values. We plot the role of the component of the dielectric tensor ($\varepsilon_z$) which is responsible for confining the evanescent waves. The increase of this constant helps to compress the evanescent waves in the cladding decreasing the mode length below the diffraction limit when the index crosses $n_z = \sqrt{\varepsilon_z} \approx 3$. We emphasize that this is within reach at optical communication wavelengths. The anisotropic metamaterial cladding also achieves a significantly better power confinement in the core as compared to the conventional waveguide (Fig. 3). This increase in power confinement is accompanied by a proportionate decrease in mode length as compared to the conventional waveguide (inset). For any given core irrespective of its subwavelength size, we can achieve sub-diffraction confinement of light and extreme power concentration in the core if the cladding anisotropy is increased. Note the diffraction limit is defined as per convention with respect to the core index where most of the power is confined. This shows that optical mode volumes can also be governed by the index felt by the evanescent fields outside of the core.

The optimum performance occurs when $\varepsilon_x \to 1^+$ and $\varepsilon_z \gg 1$. In the supplementary information we show a detailed study that the cladding anisotropy and cladding size are degrees of freedom that can be exploited to confine the mode irrespective of core index and core size.

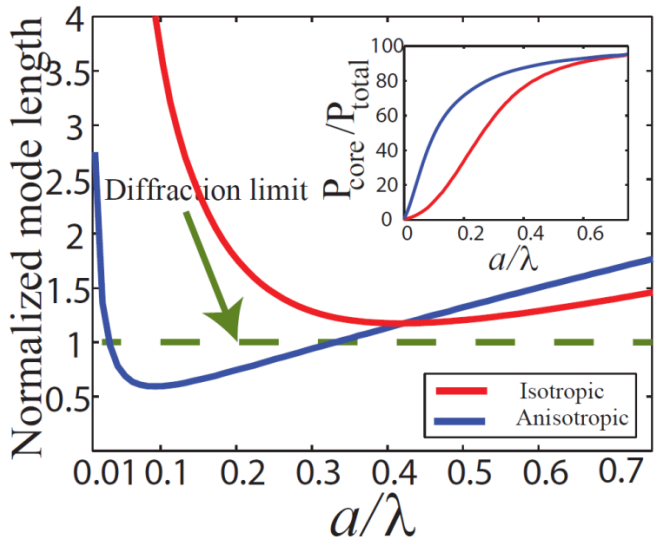

Fig. 3. Mode length comparison of slab waveguides with core size. It shows that the anisotropic cladding ($\varepsilon_x = 1.1$ and $\varepsilon_z = 15$) can confine the TM mode to sub-diffraction values. (inset) We emphasize that the net power in the core is also higher for the TCW as compared to conventional waveguides.

We have decoupled the momentum of the propagating mode in the core (related to effective mode index) from the momentum of the evanescent wave in the cladding (related to confinement). This implies that the enhanced confinement does not require a high effective mode index contrary to conventional approaches. We expect this fundamental difference to be a major design advantage for mode-matching in various devices and couplers.

**1D Practical Realization.** We discuss how to practically achieve these momentum transformations. Firstly, we argue that no naturally occurring medium has a strong anisotropy and the maximum contrast between permittivity tensor components is low for natural dielectrics (e.g. TiO$_2$) as well as artificial polymers[16]. Thus we cannot use natural dielectrics to preserve total internal reflection with a glass or silicon waveguide core interface while simultaneously increasing the momentum of evanescent waves. However, we can realize this extreme anisotropy by artificially structured media using available lossless dielectrics.

One practical approach consists of a multilayer structure consisting of two materials with a high index contrast and layer thicknesses far below the wavelength of light[17]. Effective medium theory[18] for this super-lattice predicts a homogenized medium with an anisotropic dielectric tensor given by $\varepsilon_\parallel = \varepsilon_{high}\rho + \varepsilon_{low}(1-\rho)$ where $\varepsilon_\parallel$ is the dielectric constant parallel to the layers and $1/\varepsilon_\perp = \rho/\varepsilon_{high} + (1-\rho)/\varepsilon_{low}$ is the dielectric constant perpendicular to the layers. $\rho$ is the fill fraction of the high index material $\varepsilon_{high}$.

**Application to Photonic integration** The major advantage of our approach for practical applications is the reduction in cross-talk once the metamaterial is introduced in the region between any conventional dielectric waveguides. This is because our approach relies on altering the evanescent field outside the core for confinement, the fundamental origin of cross-talk. This is a key figure of merit for photonic integration[19] and we outperform state of the art structures by an order of magnitude taking into account non-idealities.

Fig. 4.a shows schematic for two coupled slab waveguides where the cladding has been transformed to allow total internal reflection but cause fast decay of evanescent waves in the cladding to reduce the cross-talk. We consider two silicon slab waveguides ($n_{Si} = 3.47$) with a center-to-center separation of $s = 0.5\lambda$. A periodic multilayer combination of a high-index and low-index dielectric shows extreme effective anisotropy needed for the optical momentum transformation. The metamaterial claddings are made of high index ($n_1 = 4.3$) and low index thin films ($n_2 = 1.5$) at the operating wavelength of 1550 nm. Two representative materials with such indices are Germanium and Silica. We emphasize that the band edge loss at 1.55 microns in Germanium is not a fundamental impediment. For a medium 1 filling fraction of $\rho = 0.6$, multilayer effective medium theory predicts anisotropic dielectric constants of $\varepsilon_x = 4.8$ and $\varepsilon_z = 11.9$. Fig 4.b shows the coupling length of the fundamental mode in the waveguide with increasing core-size. For comparison, we also show the coupling length when the surrounding medium is simply silica. A dramatic impact of the anisotropy is clearly evident in the coupling length which shows an order of magnitude increase for various core sizes for the transformed cladding structure. For completeness, we have also compared the cross talk performance of our waveguide with another state of the art structure – slot waveguides[7,13]. Higher index cores in slot waveguides and photonic crystals can lead to enhanced power confinement however the cross-talk of our transformed cladding waveguides always outperforms them by an order of magnitude.

The anisotropic cladding can be exploited for designing bends and splitters as well. The dielectric constant of the cladding perpendicular to the core governs total internal reflection and can be minimized or altered to decrease power loss. A detailed analysis of quasi-2D waveguides, splitters and bends with extreme skin depth claddings and their performance will be presented elsewhere but we now focus on showing how our approach can be generalized to 2D waveguides.

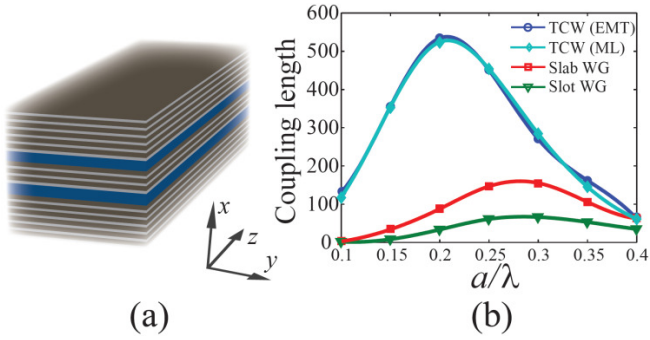

(a) (b)

Fig. 4. Dense photonic integration at optical telecommunication wavelength ($\lambda = 1550 nm$). (a) Extreme skin depth waveguide with low cross-talk between closely spaced waveguides. This can be achieved by surrounding the waveguide cores (blue) with multilayer all-dielectric metamaterials. The multilayer metamaterial consisting of alternating sub-wavelength layers Germanium (26 nm) and Silica (14 nm). This all-dielectric structure achieves the anisotropy of ($\varepsilon_x = 4.8$ and $\varepsilon_z = 11.9$) (b) Comparison of coupling length (cross-talk) for conventional slab waveguides, slot waveguides and transformed cladding waveguides. It is seen that the TCW improves the cross-talk by an order of magnitude and the practical multilayer structure result is in excellent agreement with the effectively anisotropic cladding. The core is silicon with a center to center separation of $0.5\lambda$ between waveguides. Each slot-waveguide has the same net size as the core of the other waveguides, the slot size is $0.01\lambda$ and is filled with glass. If the slot size is larger or the slot index is lower, the cross talk performance is worse than that shown above. Also note that the slot waveguide cross-talk is in fact always more than the conventional waveguide.

**2D Extreme skin depth Waveguides** The momentum transformation can be used to strongly confine light in an infinitely long glass rod with arbitrary shaped cross section ($A << \lambda^2$). The cladding has to be anisotropic to allow for the lowest-order mode (HE11) to travel inside the glass core and bounce off by total internal reflection but simultaneously decay away rapidly causing sub-diffraction confinement of the mode (Fig 5). The set of non-magnetic media which can cause the momentum transformation are anisotropic homogenous dielectric materials with $1 < \varepsilon_x = \varepsilon_y < \varepsilon_{glass}$ and $\varepsilon_z >> 1$.

The simulated electric energy density of the arbitrary shaped waveguide with an all-dielectric anisotropic cladding ($\varepsilon_x = \varepsilon_y = 1.2$ and $\varepsilon_z = 15$) is shown in fig. 5 b. The shape of the cladding is chosen to be the same shape as the arbitrary core but with twice the local radius. The numerical calculation shows that about 36% of the total power is inside the low refractive index core and the mode area for this waveguide is about $0.7A_0$ ($A_0 = (\lambda/2n_{core})^2$). Note that without the momentum transformed cladding, the fundamental mode of the subwavelength core is weakly guided and most of the power lies outside the core (fig. 5.a). The calculated mode area for the bare waveguide is about $80A_0$ and only 1% of the total power lies inside the core. Availability of high index building blocks for metamaterials at lower frequencies can lead to better performance specifically for the 2D designs[20]. Nanowire metamaterials in the cladding can achieve the desired anisotropy (see supplementary information).

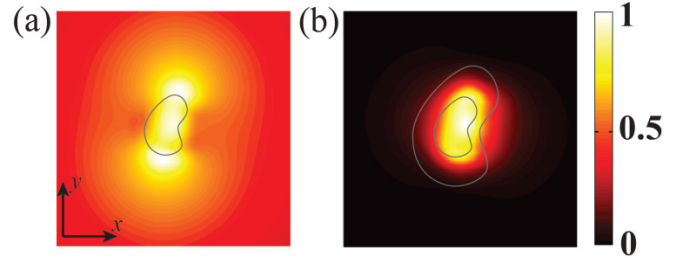

Fig. 5. Simulated distribution of the electric energy density inside a low-index 2D dielectric waveguide with arbitrary cross section using metamaterial claddings. (a) The waveguide without cladding, (b) the waveguide with all-dielectric non-magnetic cladding ($\varepsilon_x = \varepsilon_y = 1.2$ and $\varepsilon_z = 15$). When the anisotropic cladding is added, the mode area of the waveguide is decreased from about $80A_0$ to $0.7A_0$, and the fraction of power inside the core to the total power is increased from less than 1% to about 36%.

**Conclusion.** In conclusion we have introduced a paradigm shift in light confinement strategy which rests on transforming the momentum of evanescent waves. Our transformations can be achieved by all-dielectric media fundamentally overcoming the foremost challenge in the field of plasmonics and metamaterials: optical absorption. We showed that for practical device applications the introduction of our engineered anisotropy in the space between conventional waveguides confines evanescent waves and always decreases the cross talk irrespective of core index or size. The approach of altering the momentum of evanescent waves can be utilized all across the spectrum for electromagnetic waves leading to a new class of devices which work on controlling evanescent field momentum.

**Acknowledgements.** We wish to acknowledge Prashant Shekhar and Ward Newman for input.

**Funding Information**

National Science and Engineering Research Council of Canada, Canadian School of Energy and Environment, Nanobridge and University of Alberta.

See Supplement 1 for supporting content.

**References**

1. Miller, D. A. B. Device Requirements for Optical Interconnects to Silicon Chips. *Proc. IEEE* **97,** 1166–1185 (2009).
2. Ramo, S., Whinnery, J. R. & Van Duzer, T. *Fields and waves in communication electronics*. (Wiley, 1994).
3. Gramotnev, D. K. & Bozhevolnyi, S. I. Plasmonics beyond the diffraction limit. *Nat. Photonics* **4,** 83–91 (2010).
4. Han, Z. & Bozhevolnyi, S. I. Radiation guiding with surface plasmon polaritons. *Rep. Prog. Phys.* **76,** 016402 (2013).
5. Oulton, R. F., Sorger, V. J., Genov, D. A., Pile, D. F. P. & Zhang, X. A hybrid plasmonic waveguide for subwavelength confinement and long-range propagation. *Nat. Photonics* **2,** 496–500 (2008).
6. Zia, R., Selker, M. D., Catrysse, P. B. & Brongersma, M. L. Geometries and materials for subwavelength surface plasmon modes. *J. Opt. Soc. Am. A* **21,** 2442–2446 (2004).
7. Almeida, V. R., Xu, Q., Barrios, C. A. & Lipson, M. Guiding and confining light in void nanostructure. *Opt. Lett.* **29,** 1209 (2004).
8. Wiederhecker, G. S. *et al.* Field enhancement within an optical fibre with a subwavelength air core. *Nat. Photonics* **1,** 115–118 (2007).
9. Koos, C. *et al.* All-optical high-speed signal processing with silicon–organic hybrid slot waveguides. *Nat. Photonics* **3,** 216–219 (2009).
10. Joannopoulos, J. D., Villeneuve, P. R. & Fan, S. Photonic crystals: putting a new twist on light. *Nature* **386,** 143–149 (1997).
11. Krauss, T. F. Planar photonic crystal waveguide devices for integrated optics. *Phys. Status Solidi A* **197,** 688–702 (2003).
12. Lin, S.-Y., Chow, E., Hietala, V., Villeneuve, P. R. & Joannopoulos, J. D. Experimental Demonstration of Guiding and Bending of Electromagnetic Waves in a Photonic Crystal. *Science* **282,** 274–276 (1998).
13. Dai, D., Shi, Y. & He, S. Comparative study of the integration density for passive linear planar light-wave circuits based on three different kinds of nanophotonic waveguide. *Appl. Opt.* **46,** 1126 (2007).
14. Tomljenovic-Hanic, S., Martijn de Sterke, C. & Steel, M. J. Packing density of conventional waveguides and photonic crystal waveguides. *Opt. Commun.* **259,** 142–148 (2006).
15. Maier, S. A. Plasmonic field enhancement and SERS in the effective mode volume picture. *Opt. Express* **14,** 1957–1964 (2006).
16. Weber, M. F., Stover, C. A., Gilbert, L. R., Nevitt, T. J. & Ouderkirk, A. J. Giant Birefringent Optics in Multilayer Polymer Mirrors. *Science* **287,** 2451–2456 (2000).
17. Fiore, A., Berger, V., Rosencher, E., Bravetti, P. & Nagle, J. Phase matching using an isotropic nonlinear optical material. *Nature* **391,** 463–466 (1998).
18. Milton, G. W. *The Theory of Composites*. (Cambridge University Press, 2002).
19. Veronis, G. & Fan, S. Crosstalk between three-dimensional plasmonic slot waveguides. *Opt. Express* **16,** 2129 (2008).
20. Catrysse, P. B. & Fan, S. Transverse Electromagnetic Modes in Aperture Waveguides Containing a Metamaterial with Extreme Anisotropy. *Phys. Rev. Lett.* **106,** 223902 (2011).

# Transparent sub-diffraction optics: Nanoscale light confinement without metal: supplementary material

SAMAN JAHANI,[1] ZUBIN JACOB,[1,*]

[1] *Department of Electrical and Computer Engineering, University of Alberta, Edmonton, Canada T6G 2V4*
*Corresponding author: zjacob@ualberta.ca*

This document provides supplementary information to: "Transparent sub-diffraction optics: Nanoscale light confinement without metal". The full analytical mode calculations of 1D, 2D extreme-skin-depth (e-skid) waveguides, and 1D coupled waveguides are reported. The effects of waveguide parameters on power confinement, mode length/area and cross talk have been investigated. The results have been verified with CST Microwave Studio, a full-wave numerical simulation tool, and excellent agreements are observed.

**Analytical calculations**

**1D Extreme skin depth Waveguides.** For an infinite slab waveguide in which the wave propagates in the z direction, the conditions to satisfy total internal reflection and fast decaying in the cladding are: $h_z < 1$ and $h_x \gg 1$. Since for the TM modes in slab waveguides only y component of the magnetic field is non-zero, the magnetic field does not feel the anisotropy. Thus we can achieve the momentum transformation with non-magnetic anisotropic metamaterials. The magnetic field is defined as:

$$H_y = \begin{cases} A\cos(k_{x1}x)e^{i\beta z}, & |x| < a \\ B e^{-k_{x2}x}e^{i\beta z}, & |x| > a \end{cases} \quad (S1)$$

where $A$ and $B$ are constants and obtained by matching boundary conditions and excitation, and $k_{xi}$ $(i=1,2)$ and $\beta$ are the wave vector components in the perpendicular and parallel to the propagation direction, respectively. When the second medium is anisotropic dielectric, the electric field components are derived from Maxwell equations as:

$$E_x = \frac{1}{i\omega\varepsilon_0\varepsilon_x}\frac{\partial H_y}{\partial z} = \frac{\beta}{\omega\varepsilon_0}\begin{cases} \dfrac{A}{\varepsilon_1}\cos(k_{x1}x)e^{i\beta z}, & |x| < a \\ \dfrac{B}{\varepsilon_{x2}}e^{-k_{x2}x}e^{i\beta z}, & |x| > a \end{cases} \quad (S2)$$

$$E_z = \frac{-1}{i\omega\varepsilon_0\varepsilon_z}\frac{\partial H_y}{\partial x} = \frac{1}{i\omega\varepsilon_0}\begin{cases} \dfrac{Ak_{x1}}{\varepsilon_1}\sin(k_{x1}x)e^{i\beta z}, & |x| < a \\ \dfrac{Bk_{x2}}{\varepsilon_{z2}}e^{-k_{x2}x}e^{i\beta z}, & |x| > a \end{cases}$$

where $\varepsilon_1$ is the first medium permittivity and $\varepsilon_{x2}$ and $\varepsilon_{z2}$ are the x and z component of second medium permittivity. Since the structure is invariant in the y direction, other field components are zero. The isofrequency dispersion relations for the two mediums are:

$$\begin{cases} \beta^2 + k_{x1}^2 = k_0^2\varepsilon_1 \\ \beta^2 - \left(\dfrac{k_{x2}}{\gamma}\right)^2 = k_0^2\varepsilon_{x2} \end{cases} \quad (S3)$$

where $\gamma^2 = \dfrac{\varepsilon_{z2}}{\varepsilon_{x2}}$. By applying boundary conditions and combining the two equations in (S3), we obtain two coupled equations for determining wave vectors:

$$\begin{cases} w = \gamma^2 \dfrac{\varepsilon_{x2}}{\varepsilon_1} u\tan(u) \\ u^2 + \left(\dfrac{w}{\gamma}\right)^2 = (k_0 a)^2 (\varepsilon_1 - \varepsilon_{x2}) \end{cases} \quad (S4)$$

where $u = k_{x1}a$ and $w = k_{x2}a$. This set of equations can be solved graphically to find components of the wave vectors. Fig. 1 shows the graphical solution for the first even TM mode. The point where two equations in (S4) intersect is the result. The plot in fig. S1.a and S1.b are for the case in which the core is glass ($\varepsilon_1$=2.25) with size of $2a=\lambda_0/10$ and the cladding is air and anisotropic dielectric ($\varepsilon_{x2} = 1.2$ and $\varepsilon_{z2} = 20$), respectively. It is clearly observable that the anisotropic cladding leads upscaling of $w$ without a noticeable change in $u$. An increase in $w$ means faster decaying of the fields in the cladding and consequently more energy confinement inside the core. The electric and magnetic field profiles are plotted in fig. S2 in comparison with the same waveguide without cladding. It is seen that the fields decay fast in the anisotropic cladding.

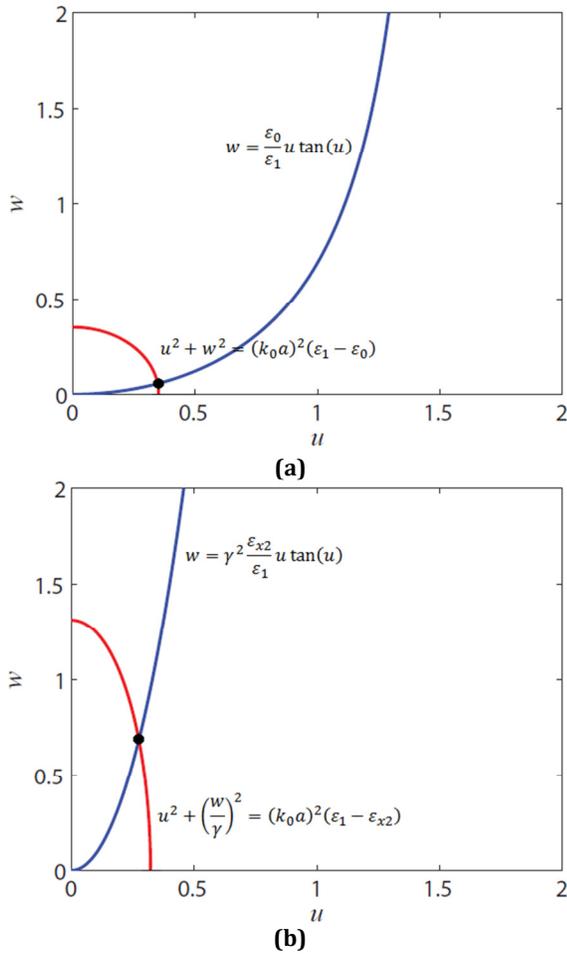

**Fig. S1.** The graphical solution of (10) for the glass slab waveguide with a size of $\lambda_0/10$ and (a) air cladding (b) anisotropic cladding ($\varepsilon_x$=1.2 and $\varepsilon_z$=20). The solution moves to higher $w$ (more confinement) without a large change in $u$.

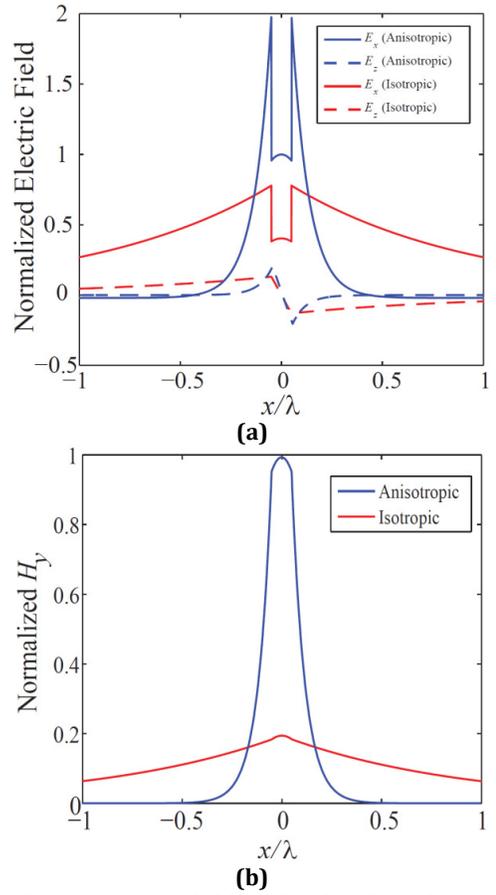

**Fig. S2.** Electromagnetic field profile for glass slab waveguide with a size of $\lambda_0/10$ and anisotropic cladding ($\varepsilon_x$=1.2 and $\varepsilon_z$=15) in comparison with the same waveguide without cladding. (a) Electric field. (b) Magnetic field. Field amplitudes are normalized to the same input energy. Fields decay faster in the anisotropic cladding.

## 2D Extreme skin depth Waveguides

**a) Dual-anisotropic cladding:** Assume an infinitely long isotropic dielectric cylinder oriented in the z direction with relative permittivity and permeability of $(\varepsilon_1, \mu_1)$ surrounded by a dual-anisotropic uniaxial dielectric with optical axis parallel to the cylinder axis and relative permittivity and permeability of $(\varepsilon_{\rho 2}=\varepsilon_{\varphi 2}=\mu_{\rho 2}=\mu_{\varphi 2}=n_{\perp 2},\ \varepsilon_{z2}=\mu_{z2}=n_{z2})$. The Helmholtz wave equation for longitudinal field components inside a uniaxial anisotropic magneto-dielectric is

$$\begin{cases} \nabla_t^2 E_z + \left(k_0^2 \varepsilon_z \mu_\perp - \dfrac{\varepsilon_z}{\varepsilon_\perp}\beta^2\right)E_z = 0 \\ \nabla_t^2 H_z + \left(k_0^2 \varepsilon_\perp \mu_z - \dfrac{\mu_z}{\mu_\perp}\beta^2\right)H_z = 0 \end{cases} \quad (S5)$$

where $\nabla_t^2$ is tangential component of Laplacian. The time-harmonic solutions of the above equations in cylindrical coordinate system for the core and cladding are

$$E_z = \begin{cases} A J_n(k_{\rho 1}\rho)\cos(n\varphi)e^{i\beta z}, & |\rho|<a \\ B K_n(k_{\rho 2}\rho)\cos(n\varphi)e^{i\beta z}, & |\rho|>a \end{cases} \quad (S6)$$

$$H_z = \begin{cases} C J_n(k_{\rho 1}\rho)\sin(n\varphi)e^{i\beta z}, & |\rho|<a \\ D K_n(k_{\rho 2}\rho)\sin(n\varphi)e^{i\beta z}, & |\rho|>a \end{cases}$$

where $A$, $B$, $C$, and $D$ are constants and defined from boundary conditions and excitations; $J_n$ and $K_n$ are $n^{th}$-order of the Bessel function of the first type and the modified Bessel function of the second type, respectively; $\beta$ is the propagation constant, and $k_{\rho 1}$, $k_{\rho 2}$ are defined as

$$\begin{cases} \beta^2 + k_{\rho 1}^2 = k_0^2 \varepsilon_1 \mu_1 \\ \beta^2 - \left(\dfrac{k_{\rho 2}}{\gamma}\right)^2 = k_0^2 n_{\perp 2}^2, \quad \left(\gamma^2 = \dfrac{\varepsilon_{z2}}{\varepsilon_{\perp 2}} = \dfrac{\mu_{z2}}{\mu_{\perp 2}} = \dfrac{n_{z2}}{n_{\perp 2}}\right) \end{cases} \quad (S7)$$

The other field components are derived from the Maxwell's equations easily:

(S8)

$$E_r = \dfrac{-1}{i\omega\varepsilon_0\varepsilon_r}\left(\dfrac{\partial H_z}{r\partial\varphi} - \dfrac{\partial H_\varphi}{\partial z}\right) = \dfrac{i}{(k_0^2\varepsilon_r\mu_\varphi - \beta^2)}\left(\beta\dfrac{\partial E_z}{\partial r} + \dfrac{\omega\mu_0\mu_\varphi}{r}\dfrac{\partial H_z}{\partial\varphi}\right)$$

$= \cos(n\varphi)e^{i\beta z} \times$

$$\begin{cases} \dfrac{i}{k_{\rho 1}^2}\left[A\beta k_{\rho 1} J'_n(k_{\rho 1}\rho) + C\dfrac{\omega\mu_0\mu_1 n}{r}J_n(k_{\rho 1}\rho)\right], & |\rho| < a \\ \dfrac{-i}{(k_{\rho 2}/\gamma)^2}\left[B\beta k_{\rho 2} K'_n(k_{\rho 2}\rho) + D\dfrac{\omega\mu_0\mu_{\perp 2} n}{r}K_n(k_{\rho 2}\rho)\right], & |\rho| > a \end{cases}$$

$$E_\varphi = \dfrac{-1}{i\omega\varepsilon_0\varepsilon_\varphi}\left(\dfrac{\partial H_r}{\partial z} - \dfrac{\partial H_z}{\partial r}\right) = \dfrac{i}{(k_0^2\varepsilon_\varphi\mu_r - \beta^2)}\left(\dfrac{\beta}{r}\dfrac{\partial E_z}{\partial\varphi} - \omega\mu_0\mu_r\dfrac{\partial H_z}{\partial r}\right)$$

$= \sin(n\varphi)e^{i\beta z} \times$

$$\begin{cases} \dfrac{-i}{k_{\rho 1}^2}\left[A\dfrac{\beta n}{r}J_n(k_{\rho 1}\rho) + C\omega\mu_0\mu_1 k_{\rho 1} J'_n(k_{\rho 1}\rho)\right], & |\rho| < a \\ \dfrac{i}{(k_{\rho 2}/\gamma)^2}\left[B\dfrac{\beta n}{r}K_n(k_{\rho 2}\rho) + D\omega\mu_0\mu_{\perp 2} k_{\rho 2} K'_n(k_{\rho 2}\rho)\right], & |\rho| > a \end{cases}$$

$$H_r = \dfrac{1}{i\omega\mu_0\mu_r}\left(\dfrac{\partial E_z}{r\partial\varphi} - \dfrac{\partial E_\varphi}{\partial z}\right) = \dfrac{i}{(k_0^2\varepsilon_\varphi\mu_r - \beta^2)}\left(\beta\dfrac{\partial H_z}{\partial r} - \dfrac{\omega\varepsilon_0\varepsilon_\varphi}{r}\dfrac{\partial E_z}{\partial\varphi}\right)$$

$= \sin(n\varphi)e^{i\beta z} \times$

$$\begin{cases} \dfrac{i}{k_{\rho 1}^2}\left[A\dfrac{\omega\varepsilon_0\varepsilon_1 n}{r}J_n(k_{\rho 1}\rho) + C\beta k_{\rho 1} J'_n(k_{\rho 1}\rho)\right], & |\rho| < a \\ \dfrac{-i}{(k_{\rho 2}/\gamma)^2}\left[B\dfrac{\omega\varepsilon_0\varepsilon_{\perp 2} n}{r}K_n(k_{\rho 2}\rho) + D\beta k_{\rho 2} K'_n(k_{\rho 2}\rho)\right], & |\rho| > a \end{cases}$$

$$H_\varphi = \dfrac{1}{i\omega\mu_0\mu_\varphi}\left(\dfrac{\partial E_r}{\partial z} - \dfrac{\partial E_z}{\partial r}\right) = \dfrac{i}{(k_0^2\varepsilon_r\mu_\varphi - \beta^2)}\left(\dfrac{\beta}{r}\dfrac{\partial H_z}{\partial\varphi} + \omega\varepsilon_0\varepsilon_r\dfrac{\partial E_z}{\partial r}\right)$$

$= \cos(n\varphi)e^{i\beta z} \times$

$$\begin{cases} \dfrac{i}{k_{\rho 1}^2}\left[A\omega\varepsilon_0\varepsilon_1 k_{\rho 1} J'_n(k_{\rho 1}\rho) + C\dfrac{\beta n}{r}J_n(k_{\rho 1}\rho)\right], & |\rho| < a \\ \dfrac{-i}{(k_{\rho 2}/\gamma)^2}\left[B\omega\varepsilon_0\varepsilon_{\perp 2} k_{\rho 2} K'_n(k_{\rho 2}\rho) + D\dfrac{\beta n}{r}K_n(k_{\rho 2}\rho)\right], & |\rho| > a \end{cases}$$

By applying the boundary conditions and subtracting dispersion relations, two coupled equations are obtained:

(S9)

$$\dfrac{J'_n(u)}{uJ_n(u)} = -\left(\dfrac{\mu_{\perp 2}}{\mu_1} + \dfrac{\varepsilon_{\perp 2}}{\varepsilon_1}\right)\dfrac{\gamma^2 K'_n(w)}{2wK_n(w)} \pm$$

$$\left[\left(\dfrac{\mu_{\perp 2}}{\mu_1} - \dfrac{\varepsilon_{\perp 2}}{\varepsilon_1}\right)^2\left(\dfrac{\gamma^2 K'_n(w)}{2wK_n(w)}\right)^2 + \dfrac{n^2\beta^2}{\varepsilon_1\mu_1 k_0^2}\left(\dfrac{1}{u^2} + \left(\dfrac{\gamma}{w}\right)^2\right)^2\right]^{\frac{1}{2}}$$

$$u^2 + \left(\dfrac{w}{\gamma}\right)^2 = (k_0 a)^2(\varepsilon_1\mu_1 - \varepsilon_{\perp 2}\mu_{\perp 2})$$

where $u = k_{\rho 1} a$ and $w = k_{\rho 2} a$. This set of equations can yield two independent solutions according to the sign of the second term in the first equation. For n>1, the solutions are conventionally known as HE and EH modes for negative and positive signs respectively, for the conventional optical fibers. The first HE mode can propagate without any cut-off.

**b) non-magnetic anisotropic cladding.** To verify that the transformed momentum of light can indeed lead to sub-diffraction confinement even without using dual anisotropic cladding, we solved Maxwell's equations in an arbitrary shaped glass core surrounded by all-dielectric nonmagnetic anisotropic metamaterials that transform momentum. Assume that the core and anisotropic cladding is nonmagnetic ($\mu_1 = \mu_2 = 1$) in (S5). The longitudinal field components in the core and cladding are:

$$E_z = \begin{cases} A J_n(k_{\rho 1}\rho)\cos(n\varphi)e^{i\beta z}, & |\rho| < a \\ B K_n(k_{\rho 2}\rho)\cos(n\varphi)e^{i\beta z}, & |\rho| > a \end{cases} \quad (S10)$$

$$H_z = \begin{cases} C J_n(k_{\rho 1}\rho)\sin(n\varphi)e^{i\beta z}, & |\rho| < a \\ D K_n\left(\dfrac{k_{\rho 2}}{\gamma}\rho\right)\sin(n\varphi)e^{i\beta z}, & |\rho| > a \end{cases}$$

where $k_{\rho 1}$, $k_{\rho 2}$ are defined as

$$\begin{cases} \beta^2 + k_{\rho 1}^2 = k_0^2 \varepsilon_1 \\ \beta^2 - \left(\dfrac{k_{\rho 2}}{\gamma}\right)^2 = k_0^2 \varepsilon_{\perp 2}, \quad \left(\gamma^2 = \dfrac{\varepsilon_{z2}}{\varepsilon_{\perp 2}}\right) \end{cases} \quad (S11)$$

The other field components are derived in the same way:

(S12)

$$E_r = \dfrac{-1}{i\omega\varepsilon_0\varepsilon_r}\left(\dfrac{\partial H_z}{r\partial\varphi} - \dfrac{\partial H_\varphi}{\partial z}\right) = \dfrac{i}{(k_0^2\varepsilon_r - \beta^2)}\left(\beta\dfrac{\partial E_z}{\partial r} + \dfrac{\omega\mu_0}{r}\dfrac{\partial H_z}{\partial\varphi}\right)$$

$= \cos(n\varphi)e^{i\beta z} \times$

$$\begin{cases} \dfrac{i}{k_{\rho 1}^2}\left[A\beta k_{\rho 1} J'_n(k_{\rho 1}\rho) + C\dfrac{\omega\mu_0 n}{r}J_n(k_{\rho 1}\rho)\right], & |\rho| < a \\ \dfrac{-i}{(k_{\rho 2}/\gamma)^2}\left[B\beta k_{\rho 2} K'_n(k_{\rho 2}\rho) + D\dfrac{\omega\mu_0 n}{r}K_n\left(\dfrac{k_{\rho 2}}{\gamma}\rho\right)\right], & |\rho| > a \end{cases}$$

$$E_\varphi = \frac{-1}{i\omega\varepsilon_0\varepsilon_\varphi}\left(\frac{\partial H_r}{\partial z}-\frac{\partial H_z}{\partial r}\right)=\frac{i}{\left(k_0^2\varepsilon_\varphi-\beta^2\right)}\left(\frac{\beta}{r}\frac{\partial E_z}{\partial \varphi}-\omega\mu_0\frac{\partial H_z}{\partial r}\right)$$

$$=\sin(n\varphi)e^{i\beta z}\times$$

$$\begin{cases}\dfrac{-i}{k_{\rho 1}^2}\left[A\dfrac{\beta n}{r}J_n(k_{\rho 1}\rho)+C\omega\mu_0 k_{\rho 1}J_n'(k_{\rho 1}\rho)\right], & |\rho|<a \\ \dfrac{i}{(k_{\rho 2}/\gamma)^2}\left[B\dfrac{\beta n}{r}K_n(k_{\rho 2}\rho)+D\omega\mu_0\dfrac{k_{\rho 2}}{\gamma}K_n'\left(\dfrac{k_{\rho 2}}{\gamma}\rho\right)\right], & |\rho|>a\end{cases}$$

$$H_r = \frac{1}{i\omega\mu_0}\left(\frac{\partial E_z}{r\partial\varphi}-\frac{\partial E_\varphi}{\partial z}\right)=\frac{i}{\left(k_0^2\varepsilon_\varphi-\beta^2\right)}\left(\beta\frac{\partial H_z}{\partial r}-\frac{\omega\varepsilon_0\varepsilon_\varphi}{r}\frac{\partial E_z}{\partial\varphi}\right)$$

$$=\sin(n\varphi)e^{i\beta z}\times$$

$$\begin{cases}\dfrac{i}{k_{\rho 1}^2}\left[A\dfrac{\omega\varepsilon_0\varepsilon_1 n}{r}J_n(k_{\rho 1}\rho)+C\beta k_{\rho 1}J_n'(k_{\rho 1}\rho)\right], & |\rho|<a \\ \dfrac{-i}{(k_{\rho 2}/\gamma)^2}\left[B\dfrac{\omega\varepsilon_0\varepsilon_{\perp 2} n}{r}K_n(k_{\rho 2}\rho)+D\beta\dfrac{k_{\rho 2}}{\gamma}K_n'\left(\dfrac{k_{\rho 2}}{\gamma}\rho\right)\right], & |\rho|>a\end{cases}$$

$$H_\varphi = \frac{1}{i\omega\mu_0}\left(\frac{\partial E_r}{\partial z}-\frac{\partial E_z}{\partial r}\right)=\frac{i}{\left(k_0^2\varepsilon_r-\beta^2\right)}\left(\frac{\beta}{r}\frac{\partial H_z}{\partial\varphi}+\omega\varepsilon_0\varepsilon_r\frac{\partial E_z}{\partial r}\right)$$

$$=\cos(n\varphi)e^{i\beta z}\times$$

$$\begin{cases}\dfrac{i}{k_{\rho 1}^2}\left[A\omega\varepsilon_0\varepsilon_1 k_{\rho 1}J_n'(k_{\rho 1}\rho)+C\dfrac{\beta n}{r}J_n(k_{\rho 1}\rho)\right], & |\rho|<a \\ \dfrac{-i}{(k_{\rho 2}/\gamma)^2}\left[B\omega\varepsilon_0\varepsilon_{\perp 2}k_{\rho 2}K_n'(k_{\rho 2}\rho)+D\dfrac{\beta n}{r}K_n\left(\dfrac{k_{\rho 2}}{\gamma}\rho\right)\right], & |\rho|>a\end{cases}$$

Similar to the dual-anisotropic case, by applying the boundary conditions, two coupled equations are obtained:

(S13)

$$\begin{cases}\dfrac{J_n'(u)}{uJ_n(u)}=-\left(\dfrac{\gamma K_n'\left(\dfrac{w}{\gamma}\right)}{2wK_n\left(\dfrac{w}{\gamma}\right)}+\dfrac{\varepsilon_{\perp 2}\gamma^2 K_n'(w)}{2\varepsilon_1 wK_n(w)}\right)\pm\\ \left[\left(\dfrac{\gamma K_n'\left(\dfrac{w}{\gamma}\right)}{2wK_n\left(\dfrac{w}{\gamma}\right)}-\dfrac{\varepsilon_{\perp 2}\gamma^2 K_n'(w)}{2\varepsilon_1 wK_n(w)}\right)^2+\dfrac{n^2\beta^2}{\varepsilon_1 k_0^2}\left(\dfrac{1}{u^2}+\left(\dfrac{\gamma}{w}\right)^2\right)^2\right]^{\frac{1}{2}}\\ u^2+\left(\dfrac{w}{\gamma}\right)^2=(k_0 a)^2(\varepsilon_1-\varepsilon_{\perp 2})\end{cases}$$

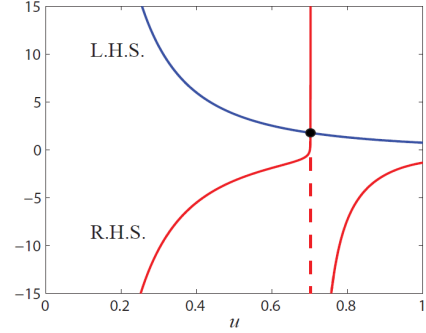

(a)

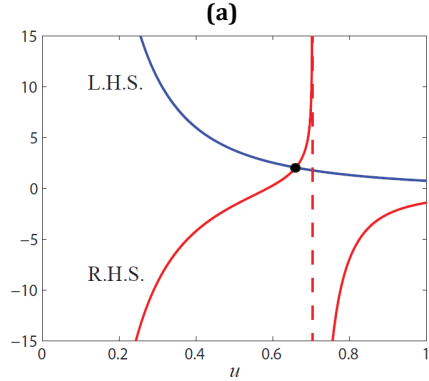

(b)

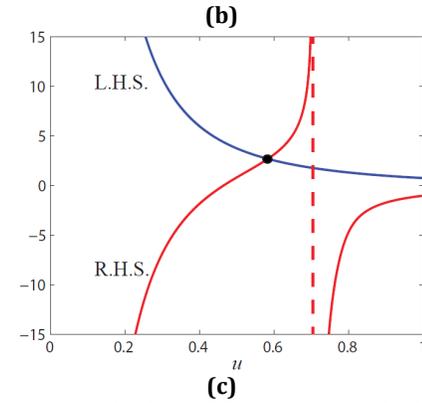

(c)

**Fig. S3.** The graphical solution by intersecting the left and right hand side of (S9) and (S13) for a cylindrical waveguide with glass core whose radius is $\lambda_0/10$ and the cladding is (a) air (b) anisotropic non-magnetic dielectric ($\varepsilon_\perp=1$ and $\varepsilon_z=20$) (c) dual-anisotropic dielectric ($\varepsilon_\perp=\mu_\perp=20$ and $\varepsilon_z=\mu_z=20$). It is seen that the waveguide mode shifts to lower values of u (higher w) denoting faster decay in the cladding.

Fig. S3 displays the graphical solution of the equation (S9) and (S13) for a cylindrical glass core with radius of $\lambda_0/10$ for air, non-magnetic anisotropic, and dual anisotropic claddings. The plots show the right and left side of the first equation in (S9) and (S13) versus *u* when *w* is replaced from the second equation in (S9) and (S13). The asymptote (shown by the dashed line) is the maximum value for the *u* where *w* becomes zero (which means that although the mode propagates without cut-off, the wave does not decay in the cladding and its size must be infinite). It can be seen that anisotropic cladding shifts the intersection to the lower values for *u* (higher *w*) without changing the asymptote position, so, anisotropic cladding leads to the faster decaying and consequently more confinement.

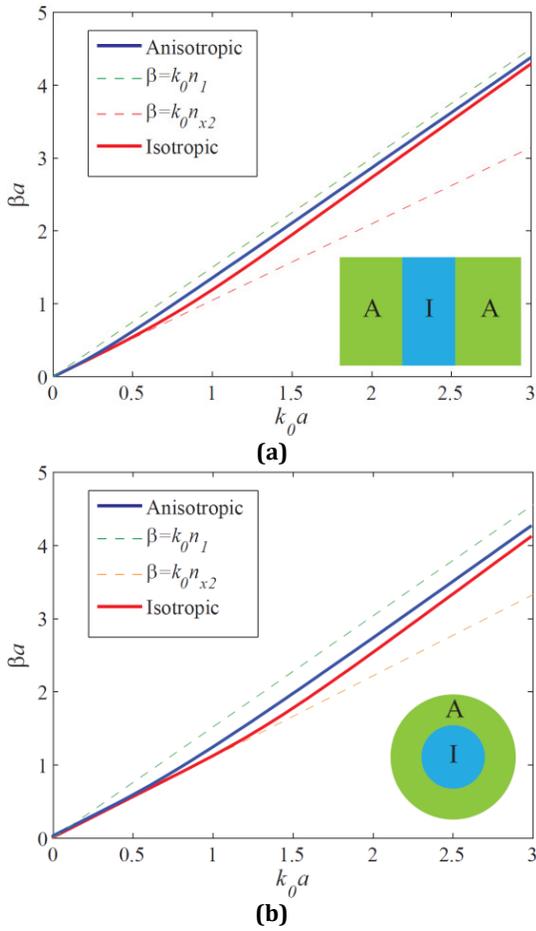

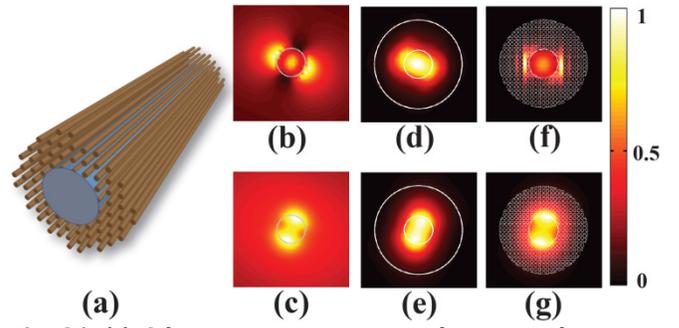

**Fig. S4.** The propagation constant of (a) 1D and (b) 2D nonmagnetic extreme skin depth (e-skid) waveguides in comparison with a conventional slab waveguide and an optical fiber with cladding permittivity of $\varepsilon = 1.1$. The core core size/radius is $\lambda_0/10$; $\varepsilon_x = 1.1$ ($\varepsilon_\perp = 1.1$) and $\varepsilon_z = 20$. The light line in the core and the cladding is also plotted, which shows the wave propagates inside the core due to the total internal reflection and decays away outside the core. The propagation constant in anisotropic case is larger since the power confinement is more.

Propagation constant dispersion for both 1D and 2D extreme skin depth (e-skid) waveguides are displayed in fig. S4. Propagation constant is always between the light lines in the core and the cladding. It guaranties that the guided mode bounces off by total internal reflection inside the core and decays away in the cladding. At low frequencies the propagation constant is close to the light line in the cladding. This means that the mode is poorly confined inside the core, but as the frequency increases, more power is confined inside the core and the propagation constant asymptotically meets the light line inside the core. Since the power confinement is better for extreme skin depth (e-skid) waveguides in comparison with conventional dielectric waveguides, the propagation constant is larger in the entire spectrum.

**Fig. S5**. (a) Schematic representation of a practical extreme skin depth (e-skid) fiber at optical telecommunication wavelength ($\lambda = 1550\,nm$). The metamaterial cladding consisting of germanium nanorods embedded in porous silica surrounding a silicon rod core. (b-g) Normalized simulated distribution of the electric (top plots) and magnetic (bottom plots) energy density of the waveguide when the core diameter is 0.15λ and germanium fill fraction is 62.5%. (b) & (c) Electric and magnetic energy density of the bare waveguide. The fraction of power inside the core to the total power (η) is 3% and the mode area is $43(\lambda/2n_{core})^2$. (d) & (e) Electric and magnetic energy density of the waveguide with anisotropic cladding ($\varepsilon_x = \varepsilon_y = 3.7$ and $\varepsilon_z = 11.8$). In the homogenized limit, the calculated η, mode area, and effective index are 35%, $1.85(\lambda/2n_{core})^2$, and 2.13, respectively. (f) & (g) Electric and magnetic energy density of the practical waveguide surrounded by nanowires which achieves the required anisotropy. According to effective theory, it achieves the effective permittivity of the cladding of part D & E. The simulated results of η, mode area, and effective index are 40%, $1.61(\lambda/2n_{core})^2$, and 2.09, respectively in agreement with the effective medium calculations. Significantly better performance can be achieved by using higher index rods in the nanowire design.

**Metamaterial fiber.** We present a practical metamaterial fiber that utilizes extreme skin depth (e-skid) cladding for enhanced confinement at the telecommunication wavelength (1550 nm). We consider a cylindrical core of silicon (n=3.47) with a diameter of 0.15λ and the cladding is composed of germanium (n=4.3) nanowires surrounded by ultra-low index porous silica [1] (n=1.05). The cladding size is three-times bigger than the core (Fig. S5.a). The germanium nanowire axis is parallel to the fiber axis and their periodicity as well as diameter is much smaller than the wavelength ($\Lambda = 20\,nm, D = 18\,nm$). Maxwell-Garnett method can be used to approximate the effective permittivity of the cladding. For the germanium filling fraction of $\rho = 0.625$, we can obtain strong anisotropy ($\varepsilon_x = \varepsilon_y = 3.7$ and $\varepsilon_z = 11.8$). It is important to note that the the dielectric constant along the propagation direction is large however the transverse dielectric constant is lesser than silicon preserving total internal reflection. A detailed theory of how the fundamental HE$_{11}$ mode of any cylindrical structure is affected by the momentum transformation is presented in the supplementary information.

Fig S5. b & c show the simulation results of the electric and magnetic energy density, respectively, for the bare waveguide. Only 3% of the total power is inside the core and the mode area is 43A$_0$. Fig. S5 d & e show the results when the anisotropic cladding is added. In these plots, we consider a homogenous effective medium cladding. In this case, the fraction of power inside the core to the total power (η) increases to 35% and the mode area decreases to 1.85A$_0$. The results for the practical

nanorod realization ($\Lambda = 20nm, D = 18nm$) are plotted in fig. S5 f & g. In this case, η=40% and the mode area is 1.61$A_0$, which is in excellent agreement with the homogenized case.

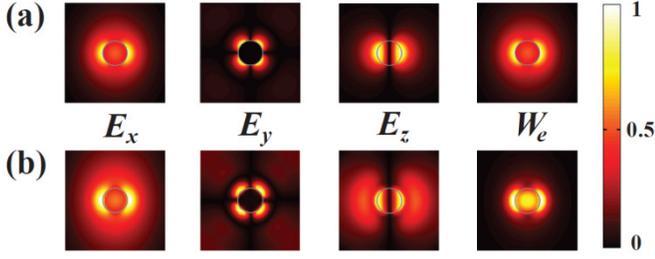

**Fig. S6.** Electric field and energy density distribution of an extreme skin depth (e-skid) waveguide composed a cylindrical silicon core with a diameter of 0.15$\lambda_0$ covered by a nonmagnetic anisotropic cladding ($\varepsilon_{\perp 2}$ =3.7 and $\varepsilon_z$ =11.8).(a) Analytical calculation. (b) Numerical calculation. The analytical calculation of η, mode area, and effective index are 34.6%, $1.94(\lambda/2n_{core})^2$, and 2.126, respectively. The simulated results are 26%, $2.58(\lambda/2n_{core})^2$, and 2.15, respectively. The slight deviations between analytical and simulated results arise due to a finite sized simulation domain.

We note that the nanowire design does not attain deeply subwavelength performance at 1.55 micron wavelength due to limitations in available high index media for the extreme skin depth (e-skid) cladding. However, we note the excellent agreement between the effective medium theory of 2D anisotropic claddings and the practical nanowire design. Therefore the cross-talk reduction over conventional designs presented in the earlier sections will hold true for the 2D design as well. We emphasize that at lower frequencies readily available high index media [2] can lead to deep subwavelength performance using both 1D multilayer and 2D nanowire designs. Comparison between analytical calculations explained above and the simulation results shown in fig. 6. There is a good agreement between the exact solution and numerical calculations. Fig. S7 shows the effect of anisotropic cladding on decaying of evanescent waves in the cladding for the structure in fig. S6.

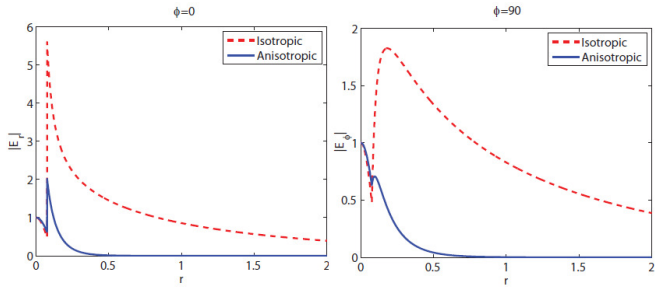

**Fig. S7**. Normalized electric field distribution of the waveguide in fig. S6 in comparison with the case without cladding. The cladding causes fields to decay fast outside in the cladding signifying the transformation of optical momentum.

### Figures of Merit

**a) Mode Length/Area:** To confirm the sub-diffraction nature of waveguide we calculated the mode area for waveguides, a common figure of merit for sub-diffraction devices [3]:

$$A_m = \frac{W_m}{max\{W(r)\}} = \frac{1}{max\{W(r)\}} \iint_{-\infty}^{\infty} W(r)d^2r \qquad (S14)$$

where $W_m$ and $W(r)$ are electromagnetic energy and energy density, respectively, and the integration is calculated on the surface normal to the propagation direction. In fig. S8 and S9. a we compare the mode length/area of the 1D and 2D extreme skin depth (e-skid) waveguides with conventional waveguides. The mode size reduction using anisotropic cladding is clearly evident.

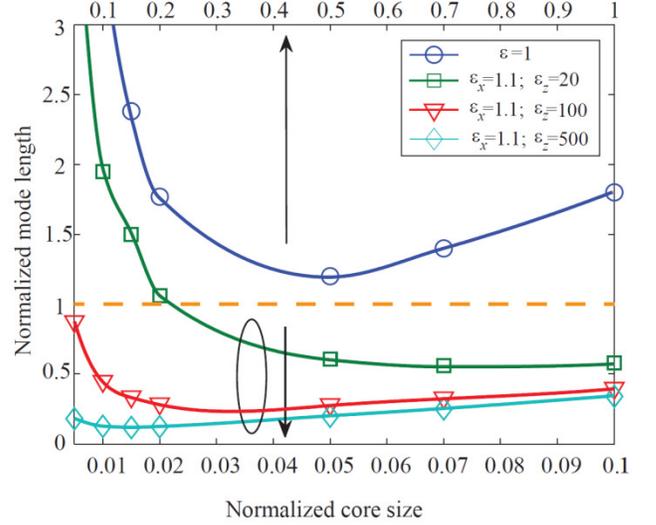

**Fig. S8.** Deep sub-diffraction performance: Irrespective of core size made of glass, one can always achieve sub-diffraction mode lengths with access to higher index media. This is difficult at optical frequencies but can be an achieved by anisotropic metamaterials at THz and microwave frequencies.

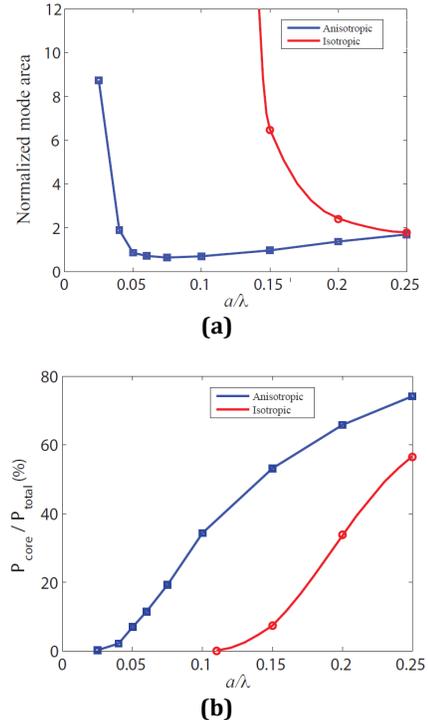

**Fig. S9.** Comparison of (a) normalized mode area and (b) power inside the core fraction for the glass 2D waveguide with and without cladding versus core radius. The cladding permittivity in tangential and longitudinal direction is 1 and 20, respectively. It is seen that the metamaterial cladding decreases

the mode area and also causes the power in the core to increase.

**b) Power in the core:** Note that the above definition of mode area essentially elucidates the peak power carried for a given input power in the waveguide. However, due to the inhomogeneous profile of the fields it becomes imperative to define the power carried within this mode area. For example, in the case of slot waveguides the evanescent tail carries a significant fraction of the power though the peak power in the slot is quite high. In fig. S9, we have evaluated both the mode area and power in the core for TCWs.

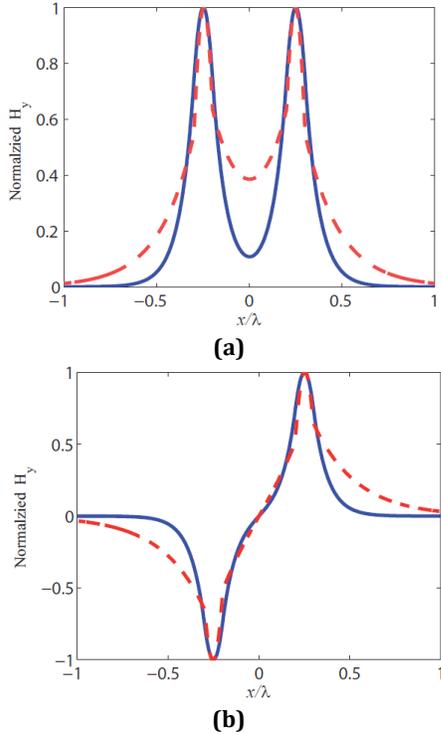

(a)

(b)

**Fig. S10.** The normalized magnetic field amplitude of the first (a) even (b) odd TM mode of coupled silicon slab waveguides with the slab size of *0.1λ* and center-to-center separation of *0.5λ* for silica (dashed red lines) and germanium-silica multilayer (solid blue lines) cladding with effective permittivity of $\varepsilon_x$ =4.8 and $\varepsilon_z$ =11.9.

**c) Cross-talk:** For dielectric waveguides it also becomes necessary to evaluate the cross talk once the size of the core decreases. This is important for understanding the potential for dense photonic integration. The major advantage of practical TCWs is the reduced cross talk.

Magnetic field of even and odd TM mode for a pair of coupled dielectric slab waveguides with size of *2a* and separation of *s* surrounded by an anisotropic dielectric can be written as:

(S15)

$$H_y^e = \begin{cases} A^e \cosh(k_{x2}^e x) e^{i\beta^e z} & |x| \leq s \\ \left[B^e \cos(k_{x1}^e x) + C^e \sin(k_{x1}^e x)\right] e^{i\beta^e z} & s < |x| \leq s+2a \\ D^e e^{-k_{x2}^e z} e^{i\beta^e z} & s+2a < |x| \end{cases}$$

$$H_y^o = \begin{cases} A^o \sinh(k_{x2}^o x) e^{i\beta^o z} & |x| \leq s \\ \left[B^o \cos(k_{x1}^o x) + C^o \sin(k_{x1}^o x)\right] e^{i\beta^o z} & s < |x| \leq s+2a \\ D^o e^{-k_{x2}^o z} e^{i\beta^o z} & s+2a < |x| \end{cases}$$

where momentums are defined similar to the single dielectric slab waveguide case shown in (S3). By matching boundary conditions and using dispersion relations, propagation constant for the even and odd modes can be calculated. We also analytically calculated propagation constant of the even and odd mode of coupled slab waveguides for half space and using symmetric and antisymmetric boundary conditions, respectively.

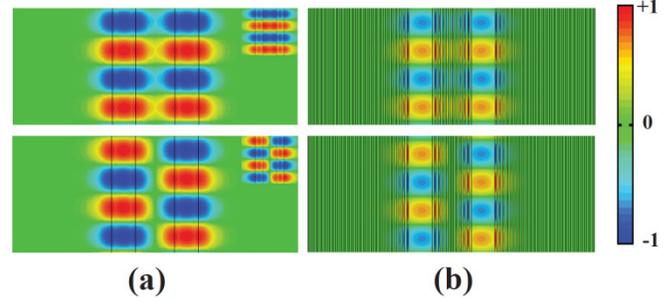

(a) (b)

**Fig. S11.** (a) Simulation results of the tangential electric field for even (up) and odd (bottom) modes of the coupled silicon waveguides. The core size is 0.1λ and center-to-center separation of 0.5λ. The required anisotropic all-dielectric response can be obtained using a germanium/silica multilayer with germanium fill fraction of 0.6. The inset shows the fields when the cladding is bulk silica. Note that the fields at the midpoint of the waveguide is significantly decreased with the TCW, a necessary condition for low cross talk. The coupling length of the structure with silica cladding is only 6λ, but it becomes more than 132λ with the metamaterial cladding. The power fraction inside the core also increases from 25% to 45%. (b) The simulation results of symmetric (up) and antisymmetric (bottom) modes for the practical structure with germanium/silica multilayer cladding and a unit cell size of 40 nm. The fields show good correspondence with the effective medium approximation and the coupling length is 119λ.

Fig. S10 compares the normalized magnetic field of the even and odd mode of two coupled silicon slab waveguides with a size of λ/10 and the separation of λ/2 when the cladding is air (dashed lines) and when it is anisotropic (solid lines). In this case, the anisotropic dielectric permittivity is derived from effective medium theory (EMT) for silica-germanium multilayer at optical telecommunication wavelength (1550 nm) when the germanium fill fraction is 0.6. The EMT formulations for multilayers and nanorods are presented in the next section. The calculated propagation constant for the even and odd mode is $1.73k_0$ and $1.64k_0$, so the coupling length ( $L_c \equiv \pi / |\beta^e - \beta^o|$ ) for silica cladding is less than 6λ, but when the waveguides are covered by the anisotropic cladding, the propagation constant of the two modes become closer ( $\beta^e = 2.5490k_0$ $\beta^o = 2.5448k_0$ ) and the coupling length increases dramatically and becomes 119λ. For the practical multilayer cladding, $\beta^e = 2.5646k_0$ and $\beta^o = 2.5605k_0$ ascertained through full wave numerical simulations shown in fig. S10.

In fact, for large coupling length, the propagation constant and consequently fields amplitude of the even and odd modes should be very close. Note that the magnetic field for odd mode is zero at the center. The coupling length increases if the wave decays fast outside the cores and become negligible at the center for the even mode. As we can see in fig. 10 and 11, the momentum transformation helps to reduce field amplitude at the center and decrease crosstalk.

We have also simulated the coupling length versus the cladding size shown in fig. S12. It is seen that the performance approaches the ideal infinite cladding for sizes approximately 3 times the core. The effect of core index on power confinement and crosstalk is displayed in fig. S13. As the core index increases, the effect of the anisotropic cladding on power confinement is modest but the cross talk in extreme skin depth (e-skid) waveguides is almost two orders of magnitude less than that in conventional waveguides.

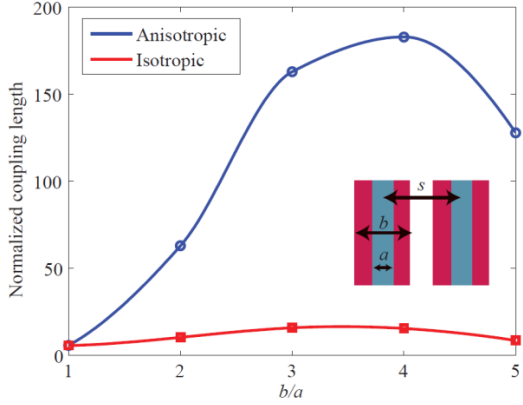

**Fig. S12.** Normalized coupling length versus the cladding size. The core is silicon (n=3.47) with a size of 0.1λ and center to center separation of 0.5 λ. When the cladding is anisotropic ($\varepsilon_x = 4.8$ and $\varepsilon_z = 11.9$) the coupling length is much larger than that when the cladding is isotropic ($\varepsilon = 4.8$). The waveguides are surrounded by glass.

**Practical realization 1D and 2D**. We discuss how to practically achieve these momentum transformations. Firstly, we argue that no naturally occurring medium has a strong anisotropy and the the maximum contrast between permittivity tensor elements is low for natural dielectrics (e.g. TiO$_2$) as well as artificial polymers [8]. Thus we cannot use natural dielectrics to preserve total internal reflection with a glass or silicon waveguide core interface while simultaneously increasing the momentum of evanescent waves. However, we can realize this extreme anisotropy by artificially structured media using available lossless dielectrics.

One practical way for realization of extreme anisotropic metamaterials is embedding periodic thin high index nanorods in a low index host dielectric such that the periodicity is much lower than the operating wavelength to ensure that the structure behaves as a homogenous material and it is far away from its band-gap. Effective relative permittivity normal ($\varepsilon_\perp$) and parallel ($\varepsilon_\parallel$) to nanorods axis for subwavelength conditions is independent to the periodicity and is calculated using Maxwell-Garnett approximation as [4]:

$$\varepsilon_\parallel = \rho \varepsilon_d + (1-\rho)\varepsilon_h \quad \text{(S16)}$$

$$\varepsilon_\perp = \frac{(1+\rho)\varepsilon_d \varepsilon_h + (1-\rho)\varepsilon_h^2}{(1-\rho)\varepsilon_d + (1+\rho)\varepsilon_h}$$

where $\varepsilon_d$ and $\varepsilon_h$ are permittivity of the dielectric nanorods and host, respectively, and $\rho$ is the fill-fraction of a nanorods in its unit-cell. According to (S16), thin high index nanorods in a low-index host behave effectively as extreme anisotropic material to transform the desired momentum transformation in the cladding of the proposed cylindrical waveguides.

Another structure for realization of extreme anisotropic metamaterials is multilayer combination of a high index and low index dielectric. In this case, effective relative permittivity normal and parallel to the layers are calculated as [4]

$$\varepsilon_\parallel = \rho \varepsilon_d + (1-\rho)\varepsilon_h \quad \text{(S17)}$$

$$\varepsilon_\perp = \frac{\varepsilon_d \varepsilon_h}{(1-\rho)\varepsilon_d + \rho \varepsilon_h}$$

This structure can be used for the momentum transformation in single and coupled slab waveguides as well as silicon strip waveguides.

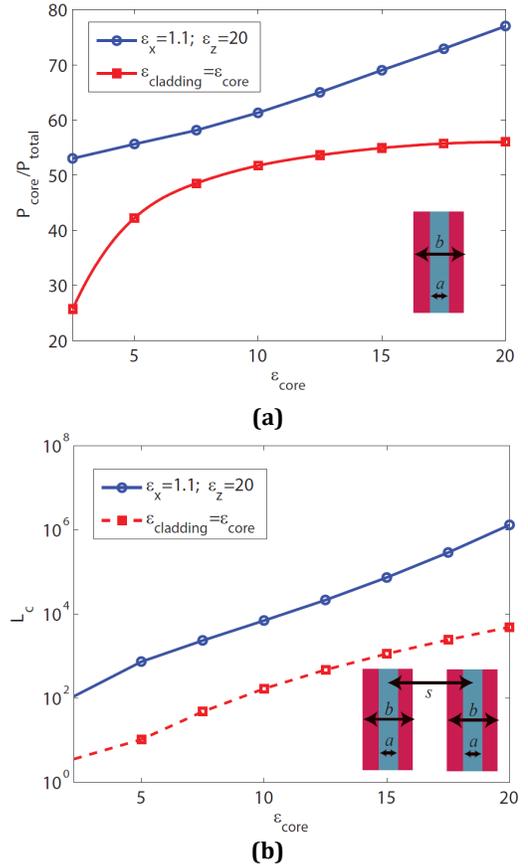

**Fig. S13.** Role of core index, core size, cladding anisotropy and cladding size. (a) Comparing power confinement in the slab waveguide with anisotropic cladding ($\varepsilon_x = 1.1$ and $\varepsilon_z = 20$) as opposed to isotropic cladding. The core size is $a$=0.1λ and the cladding size is three-times bigger than the core. It is seen that the power confinement in the core can be larger for the TCW even for high index cores. (b) Coupling length in the slab waveguide with anisotropic cladding ($\varepsilon_x = 1.1$ and $\varepsilon_z = 20$) is two orders of magnitude larger when compared to the isotropic cladding. The center-to-center separation of the coupled waveguides is $s$=0.5λ. The cladding size is three-times bigger than the core surrounded by air. The core size is $a$=0.1λ.

**Comparison of different waveguide classes**. The main figures of merit for a nano-waveguide is the length of propagation, confinement (mode area), power in the waveguide core (or within the mode area) and cross-talk between waveguides to assess the possibility of dense integration. Now we compare these figures of merit for the most well-known optical waveguides:

*Optical fibers:* It is almost loss-less and is appropriate for long distances communications, but it is diffraction-limited so energy cannot be confined for sub-diffraction sizes. Moreover, the wave decays slowly inside the cladding thus it is not suitable for photonic integration.

*Photonic crystal fibers:* It works based on Bragg reflection not total internal reflection. Thus the power does not scatter at bends and power can be confined inside a low index core. However, when the core size and period numbers becomes very small, power leaks outside so photonic integration density and power confinement is limited [5].

*Plasmonic waveguides:* Metals, in contrast to dielectrics, can confine energy below diffraction limit of light, but metals are too lossy at optical frequencies and even very low-loss metallic waveguides [3] cannot propagate light more than several wavelengths.

*Slot waveguides:* In this type of dielectric waveguides [6], light can be confined in a low-index sub-diffraction slot surrounded by a high-index dielectric. This kind of waveguide is suitable for nonlinear applications because of its small mode area, but since only a small fraction of power is inside the core (in comparison with optical fibers with similar integrated density) cross-talk for slot-waveguides is higher [7], so they are not suitable for confined photonic circuits.

*Extreme skin depth (e-skid) waveguides:* On the other hand, by transforming the space around a low-index dielectric, it is possible to confine energy inside a sub-diffraction core. Since both peak of energy density inside the core and decaying in cladding increases, photonic integration density increases and it is also suitable for nonlinear processes.

**Methods**. We have used CST Microwave Studio [9], full-wave commercial software based on a finite integration technique (FIT), to obtain simulation results for the coupled slab waveguides reported in fig. 4, for the waveguides with arbitrary cross section in figs. 5, and for the metamaterial fibers in supplementary information. Hexahedral meshes with 40 lines per wavelength with lower mesh limit of 10 have been assigned for plots in fig. 4, and 5. For plots in fig. 5 the time-domain solver has been used. The simulation area in transverse plane has been $2\lambda \times 2\lambda$ for the waveguides with anisotropic cladding and $10\lambda \times 10\lambda$ for the bare waveguides. The mode area and power ratio have been estimated based on the template-based post-processing 2D integral calculations. For plots in fig. 4, we have used the frequency domain solver. The simulation domain for plots in fig. 4 length in y direction is only $\lambda/40$, terminated to magnetic boundary conditions to model infinite width for the slab waveguides. The open boundaries in x directions have been assigned one $\lambda$ away from slab waveguides. The propagation constants have been derived from port information 1D results, and are in good agreement with analytical calculations. The simulations were made to converge with a maximum residual energy inside the calculation domain of $10^{-6}$ and $10^{-4}$ for time domain and frequency domain simulations, respectively.

## References

[1]. Xi, J.-Q. *et al.* Optical thin-film materials with low refractive index for broadband elimination of Fresnel reflection. *Nature Photonics* **1,** 176–179 (2007).

[2]. Catrysse, P. B. & Fan, S. Transverse Electromagnetic Modes in Aperture Waveguides Containing a Metamaterial with Extreme Anisotropy. *Phys. Rev. Lett.* **106,** 223902 (2011).

[3]. Oulton, R. F., Sorger, V. J., Genov, D. A., Pile, D. F. P. & Zhang, X. A hybrid plasmonic waveguide for sub-wavelength confinement and long-range propagation. *Nature Photon.* **2**, 495–500 (2008).

[4]. Milton, G. W. *The Theory of Composites*. (Cambridge University Press: 2002).

[5]. Marpaung, D. *et al.* Integrated microwave photonics. *Laser & Photonics Reviews* 1–33 (2013). doi:10.1002/lpor.201200032

[6]. Almeida, V. R., Xu, Q., Barios, C. A. & Lipson, M. Guiding and confining light in void nanostructure. Opt. Lett. **29**, 1209–1211 (2004).

[7]. D. Dai, Y. Shi, and S. He, "Comparative study of the integration density for passive linear planar lightwave circuits based on three different kinds of nanophotonic waveguide," *Appl. Opt.* **46**(7), 1126–1131 (2007).

[8]. Weber, M. F., Stover, C. A., Gilbert, L. R., Nevitt, T. J. & Ouderkirk, A. J. Giant Birefringent Optics in Multilayer Polymer Mirrors. *Science* **287,** 2451–2456 (2000).

[9]. *CST Studio Suite*. at <http://www.cst.com/>